# A simple, space constrained NIRIM type reactor for chemical vapour deposition of diamond

Evan L.H. Thomas[*,a], Laia Ginés[*,b], Soumen Mandal, Georgina M. Klemencic, Oliver A. Williams

School of Physics and Astronomy, Cardiff University, Queen's Buildings, The Parade, Cardiff, CF24 3AA, United Kingdom

[*] The authors contributed equally to the work.

[a] elhthomas@gmail.com.

[b] GinesL@cardiff.ac.uk.

**Abstract**

In this paper the design of a simple, space constrained chemical vapour deposition reactor for diamond growth is detailed. Based on the design by NIRIM, the reactor is composed of a quartz discharge tube placed within a 2.45 GHz waveguide to create the conditions required for metastable growth of diamond. Utilising largely off-the-shelf components and a modular design, the reactor allows for easy modification, repair, and cleaning between growth runs. The elements of the reactor design are laid out with the CAD files, parts list, and control files made easily available to enable replication. Finally, the quality of nanocrystalline diamond films produced are studied with SEM and Raman spectroscopy, with the observation of clear faceting and a large diamond fraction suggesting the design offers deposition of diamond with minimal complexity.

## I. INTRODUCTION

The outstanding and unique properties of diamond[1] have led to the pursuit of new methods to synthesise the material for both industrial and academic purposes[2]. Almost concurrently in the 1950s the two vastly different techniques of high-pressure high-temperature (HPHT)[3] and chemical vapour deposition (CVD) were pioneered. Doing away with the extreme conditions required during HPHT, the low pressure, kinetically governed growth process of CVD was successfully developed by Eversole and independently by *Angus et al.* and *Deryagin et al.*[4]

Unreported till the early 1960s, the first work published on the technique by Eversole demonstrated the successful deposition of diamond on natural diamond powder under less than atmospheric pressure using thermally decomposed hydrocarbons or carbon monoxide[5,6,7]. The subsequent introduction of hydrogen in the gas phase in 1966[8] as an etchant of non-diamond $sp^2$ carbon, first as molecular hydrogen[9] and then disassociated through either the use of a hot filament[10] or plasma activation[11], then made CVD a viable method for the growth of high quality diamond.

After two decades of the technique remaining largely unchanged, work performed at National Institute for Research in Inorganic Materials (NIRIM) in Japan in the 1980s brought about significant advances in the CVD process and reactor design[12]. Studies published by the group demonstrated growth on non-diamond substrates using a mixture of hydrogen and methane, initially ionised with a hot-filament[10] and shortly after within a plasma[11], with sufficient detail such that the methods used could be replicated by others. In particular the microwave plasma CVD reactor proposed, now termed the NIRIM reactor, offered simple operation and the prospect of high quality films free from filamentary impurities. With the design subsequently made commercially available by New Japan Radio, the reactor brought microwave plasma CVD within the reach of those interested in academia and industry.

The reactor overview proposed by the group consists of a quartz vacuum tube fed with hydrogen and methane gaseous precursors, placed through a 2.45 GHz waveguide. Through careful placement of the substrate within the waveguide and the use of a sliding short, the sinusoidal variation in electric field of the $TE_{10}$ mode along the width of the cavity results in the formation of a plasma within the centre of the cavity suitable for diamond growth[13]. While still routinely used for high quality diamond growth[14], the constraints placed on the plasma by the dimensions of the waveguide resulted in the end of dominance of the design. Commercially available reactors targeted towards research and development applications then moved towards designs coupling the $TE_{10}$ waveguide to the $TE_{01}$ mode of a cylindrical cavity. However, the increased deposition areas attainable came with both an increase in complexity of the design and the use of non-standard parts, resulting in a significant barrier to entry into the field. As such, researchers were dissuaded from investing in the microwave plasma CVD technique and growth capability limited to a select few.

This manuscript therefore details the design of a simple NIRIM type reactor, largely constructed from off-the-shelf microwave and vacuum components, as a simple and economic alternative to commercial systems, with sufficient detail to allow replication. The design lends itself well to both doped and intrinsic growth, with any deposition on the reactor walls largely present on an easily interchangeable quartz tube comprising the majority of the discharge tube. The pertinent aspects of the design and peripheral items used are detailed, while the CAD files, parts list, and control program of the reactor described below are all freely available to download and modify, all with the aim of expanding diamond growth capability[15].

## II. SYSTEM OVERVIEW

### A. Reactor Design

Based upon the design by NIRIM[11], the reactor consists of a quartz vacuum tube placed within the anti-node of a 2.45 GHz rectangular waveguide to create a plasma discharge suitable for CVD of diamond. To minimise complexity all microwave components are within the WR340 series by Sairem. The vacuum tube has been designed around an easily replaceable quartz tube and predominantly relies on readily available KF fittings. Figure 1 gives an overview of the complete reactor within panel (c), while panels (b) and (a) detail the pertinent microwave components and show a cross section of the discharge tube within the waveguide respectively. Table 1 meanwhile itemises the parts highlighted in the CAD diagrams.

Referring to Figure 1, the reactor is built around a Sairem supplied downstream source composed of a water cooled cavity, 'A', and two interchangeable chimneys to the top and bottom, 'B'. The discharge tube is then placed within the waveguide through feeding through two holes to the top and bottom of the metallic cavity, with the protruding sections of the tube covered by the metallic chimneys to prevent microwave leakage. A compressed spring running around the circumference of the inner wall of each chimney in part provides the mechanical support required to keep the discharge tube in place. The sample is meanwhile sat in a 1 inch sample holder, 'C', positioned on top of a quartz rod within the quartz tube, as shown within the cross sectional detail of panel (a). The hydrogen and methane gaseous precursors are fed into the quartz tube through the KF40 x KF25 reducing cross to the top of the reactor, 'D', from a series of MKS πMFC Mass Flow Controllers (MFCs) and pneumatic solenoid valves.

The pressure within the chamber is then monitored through the use of a Vacuubrand capacitive VSK3000 pressure gauge, and regulated with a variable speed Vacuubrand MV 2 NT VARIO diaphragm pump and CVC 3000 controller connected to one of the lower KF flanges to ensure gas movement through the chamber, 'E'. To prevent over pressurisation of the quartz tube a MKS 51A baratron pressure switch with a setpoint of 70 Torr and an above atmosphere Pfeiffer AVA 016 X pressure relief valve are connected to a KF25 cross, 'F', along with a manual diaphragm valve to allow venting to atmospheric pressure. Finally a Williamson DWF-24-36C dual wavelength pyrometer is attached to a custom made tilt stage with vacuum tight quartz window, 'G', situated atop the reducing tee to monitor the temperature of the sample during growth.

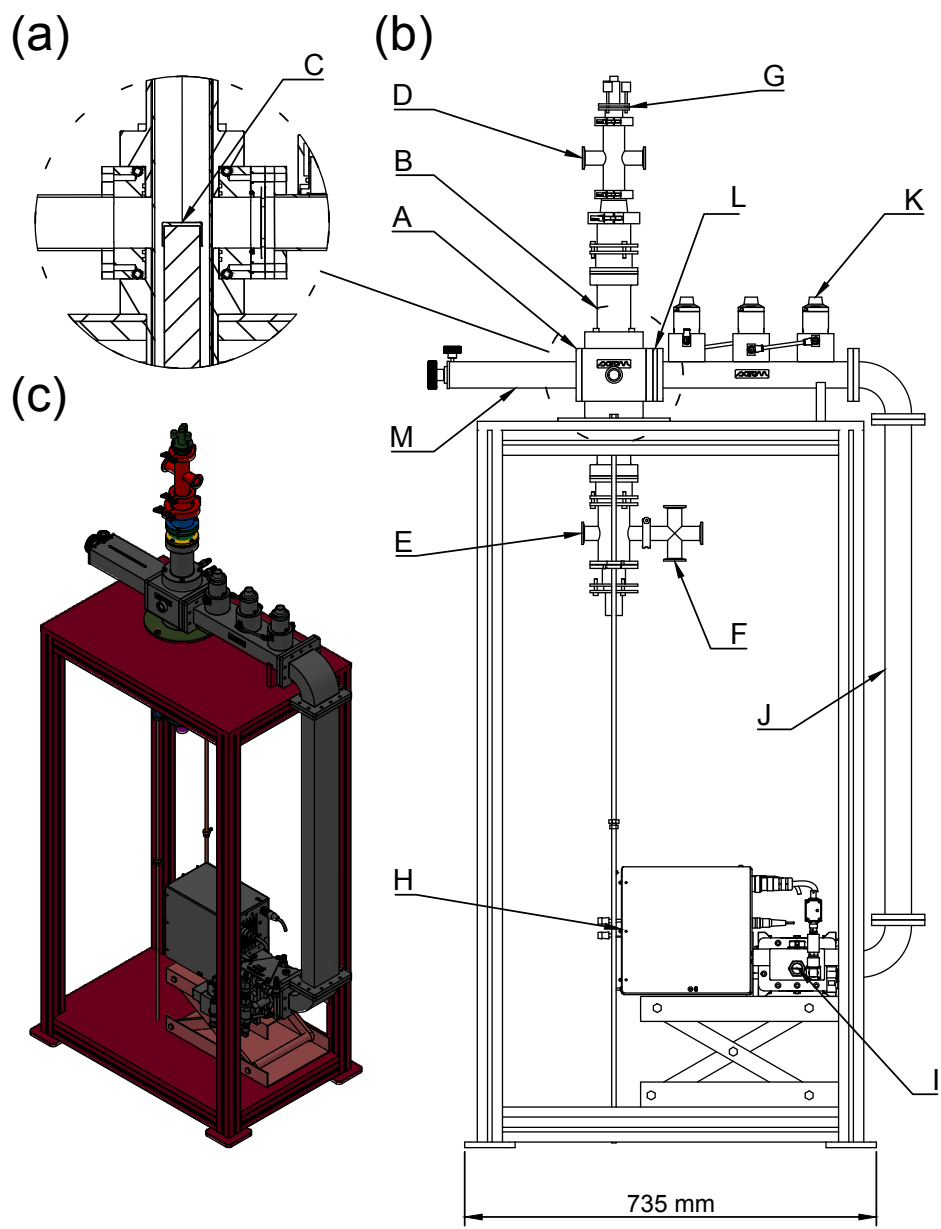

*Figure 1. NIRIM reactor overview. (a) Cross section detailing the placement of the sample holder within the middle of the downstream source, (b) side view of the reactor laying out the pertinent components, and (c) isometric view of the reactor and support structure. The itemised parts are laid out in Table 1.*

|   |   |
|---|---|
| A | Water cooled cavity |
| B | Interchangeable chimney |
| C | 1" ceramic sample holder |
| D | Off-the-shelf KF40 x KF25 reducing cross |
| E | KF25 flange for mating to standard vacuum components |
| F | Off-the-shelf KF25 cross |
| G | Pyrometer gimble and quartz window |
| H | 2 kW 2.45 GHz magnetron |
| I | Isolator and arc detection flange |
| J | WR340 waveguide |
| K | Tuning stubs |
| L | Quartz window |
| M | Sliding short-circuit |
| N | 50 mm O.D. x 46 mm I.D. quartz tube |
| O | Quartz sealing O-ring |
| P | Upper tube end |
| Q | Lower tube end |
| R | M5 tightening screws |
| S | Grooved inner alignment O-ring |
| T | 1" ceramic sample holder |
| U | Quartz rod |
| V | Rod sealing and alignment O-rings |
| W | 8 mm guide rails |
| X | Loading flange sealing O-ring |

*Table 1. Parts highlighted within Figures 1 and 2.*

Radiation is generated by a 2 kW 2.45 GHz Sairem magnetron, 'H', positioned within the support structure to minimise the footprint of the reactor. The microwaves are then fed through an isolator and arc detection flange, 'I', to minimise damage to the magnetron from both power reflected from the vacuum chamber and unintended breakdowns in the vicinity of the head before entering a $TE_{10}$ waveguide, 'J'. After passing through two 90° E-bends to the topside of the support structure, a series of tuning stubs, 'K', are used to match the impedance of the discharge tube to the magnetron source to ensure the power is delivered effectively with minimal reflection. The microwaves then finally pass through a quartz window, 'L', to prevent the ingress of dust into the waveguide and any associated damage to the microwave components to the bottom of the support structure. Upon entering the downstream source holding the quartz discharge tube, the microwaves ionise the hydrogen and methane gaseous precursors creating the conditions necessary for diamond growth. A sliding short-circuit, 'M', positioned approximately a quarter of a wavelength from the centre of the discharge tube is used to manually adjust the location of the electric field maximum within the downstream source. As such, the plasma is then positioned to coincide with the centre of the sample within the waveguide, as shown within the cross section of Figure 1(a), to realise high quality diamond growth. To simplify operation, the section of waveguide between the quartz window and the magnetron are kept at atmospheric pressure with all process gas contained within the quartz tube. The downstream source, and inadvertently the sliding short as a result of being adjacent, are meanwhile fed with compressed air to prevent overheating of the quartz tube.

Figure 2 then provides more detail on quartz tube and sample rod, with panel (a) giving an overview of the vacuum chamber with the quartz tube sectioned, and panels (b) and (c) showing cross sections of the chamber and sample loading mechanism respectively. Itemisation of the parts laid out is continued in Table 1. As shown in 2(b), the vacuum chamber is based around a 50 mm O.D. x 46 mm I.D. x 480 mm long GE214A fused quartz tube, 'N'. At either end of the tube a BS1806 50.17 mm ID x 5.33 mm thick Viton O-ring, 'O', is compressed against the quartz tube to mate to two custom made 304 stainless steel tube ends, 'P' and 'Q', by the tightening of three M5 screws, 'R'. The screws are tightened a quarter of a turn at a time while using a Vernier scale to ensure the separation between the 'clamp' and 'sheath' that constitute each tube end is uniform around the tube. Alignment of the axis of the components to that of the quartz tube is maintained through the use of a further inner O-ring, 'S', ensuring flush contact with the chimneys either side of the downstream source to minimise microwave leakage. Deviations in the O.D. of the quartz tube from the nominal 50 mm are tolerable due to the flexibility of the support springs and Viton O-rings, and the Sairem supplied components and tube ends all possessing an inner diameter of 52 mm. To prevent the formation of a trapped volume between these two pairs and allow leak testing of the reactor the inner O-rings are grooved to allow gas to freely pass. With the quartz tube constituting the majority of the length of the vacuum tube and extending far beyond the plasma, absorbates and film deposition on the stainless steel components are minimised. The reactor can then be easily kept clean through replacement of the economical quartz tube, which also aids the prevention of cross-contamination upon switching dopants within the feed gas.

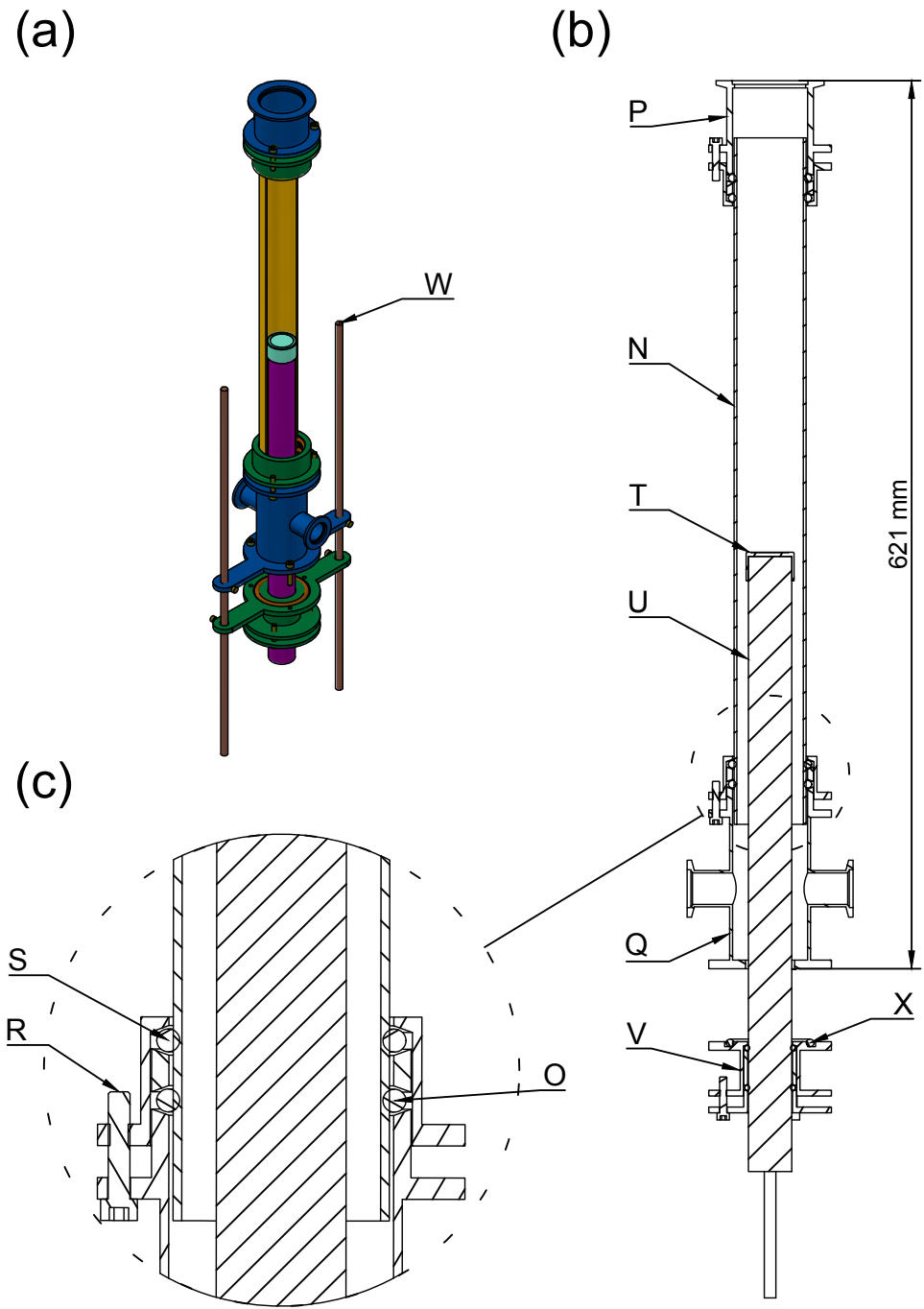

*Figure 2. Vacuum discharge tube. (a) Overview of the discharge tube with quartz tube sectioned, (b) section view of the discharge tube laying out the pertinent components, and (c) detail of the sealing and alignment mechanism around the quartz-tube. The itemised parts are laid out in Table 1.*

A Shapel M aluminum nitride sample holder, 'T', sits on top of a 30 mm OD quartz rod, 'U'. In a similar fashion to the quartz tube, a series of BS1806 28.17 mm ID x 3.53 mm thick Viton

O-rings seal the loading flange, 'V', against the quartz rod. To facilitate rapid sample loading, the loading flange slides along a pair of 8 mm guide rails bolted to the top and bottom of support structure, 'W'. Upon coming into contact with the flange on the bottom of the lower tube end and tightening with a further three M5 bolts an inset 50.17 mm I.D. x 5.33 mm thick Viton O-ring is compressed to seal the joint, 'X'. Finally, the upper and lower tube ends are terminated by a KF50 and two KF25 flanges respectively to allow connection to the KF fittings detailed previously. Once complete, mechanical support of the quartz tube is then facilitated through the combination of the uppermost tube end sitting flush against the upper chimney, the compressed springs within the chimneys, and finally the bolting of the lower tube end to the guide rails.

**B. Reactor Control and Safety Implementation**

To facilitate safe and automated operation of the reactor the MFCs and CVC 300 pump controller are interfaced with Labview through a series of National Instruments USB-6001 multifunction I/O DAQs and a USB-232 serial instrument controller. The virtual instrument (VI) created then allows dynamic control and recording of the feed gas flow rates while preventing the selection of combustible hydrogen/oxygen mixtures, and pressure within the discharge tube. Finally, the temperature of the sample is recorded through passing the 4-20 mA signal of the Williamson pyrometer remote interface through a 500 $\Omega$ resistor and measuring the potential with the inputs of the DAQs.

Should the flow rate of a MFC deviate from its set point for an extended period of time the VI will shut off all MFCs and evacuate the discharge tube. Once the pressure within the tube is no longer able to sustain a plasma, the large increase in power reflected towards to the magnetron will be promptly detected by the Sairem components, and the magnetron switched off. Similarly, should the pressure within the reactor inadvertently exceed 70 Torr the hardwired Baratron pressure switch will shut off the gas supply through the pneumatic solenoid valves placed in between the MFCs and reactor inlet. Finally, should the water flow to the reactor and microwave components fall below 6 L/min a flow switch will shut off the magnetron to prevent overheating of the reactor components. As such the reactor is designed to operate safely and without supervision during long growth runs, with processes in place to gracefully shut down should problems arise.

## III. REACTOR USAGE

### A. Film Growth

To test the capability of the design and the quality of resulting growth a series of nanocrystalline diamond (NCD) films were grown with the constructed reactor. Silicon (100) one-inch wafers were used as substrates and cleaned by RCA standard clean 1 (SC-1)[16] before use. To obtain the high nucleation densities required for the formation of coalesced thin films[17] the substrates were first seeded in a mono-dispersed colloid containing 5 nm hydrogen-terminated nanodiamond seeds/DI water, as explained by *Williams et al*[18]. The substrates were then placed in the ceramic holder sitting on top of the quartz rod previously described ('T' within Figure 2(b)), and introduced through the bottom of the reactor, as shown in Figure 1. The amount that the rod protrudes into the vacuum chamber is then adjusted such that the sample sits just below the entry to the waveguide ensuring the plasma is atop the holder and, as heating of the sample arises solely from the plasma, that the sample temperature is conducive to diamond growth. Previous work conducted on the effect of feed gas composition has indicated a reduction in grain size and a corresponding increase in non-diamond content present within films upon increasing the methane admixture[19], with these changes expected to result in a degradation of the mechanical[19], optical[20] and thermal properties[21] of resulting films. Thus the flow rates of the methane and hydrogen precursors were initially set to 15 SCCM and 285 SCCM respectively (5% $CH_4/H_2$) to establish the nanodiamond seed particles before subsequently reducing the methane content after 3 minutes to 3 SCCM (1% $CH_4/H_2$). A corresponding increase in the hydrogen flow rate to 297 SCCM ensured that a constant gas flow was maintained. Similarly, the use of low pressures[22] and low power density conditions[19] have shown to yield similar reductions in the grain size. Therefore the pressure was kept at 55 Torr while the microwave power was maintained at 0.52 kW, ensuring a large enough high-density plasma to maximise growth uniformity while minimising contact with the quartz discharge tube. The substrate temperature, as measured by dual wavelength pyrometry, was determined to be 740 °C. The resulting diamond films were grown to 1μm thick at growth rates of 0.5 μm/hour. After the termination of growth, the samples were cooled down in a purely $H_2$ fed plasma to prevent non-$sp^3$ carbon formation[23]. The morphology of the resulting samples was characterized with a Raith e-Line Scanning Electron Microscope (SEM), operating at 20 kV and a working distance of 10 mm, with images of one such film shown in Figure 3. As visible from the high magnification image in panel (b) the diamond grains exhibit clear faceting and a narrow distribution in size around 100 nm. The lower magnification image in panel (a) meanwhile shows clear coalescence of the film with a lack of pinholes to the substrate.

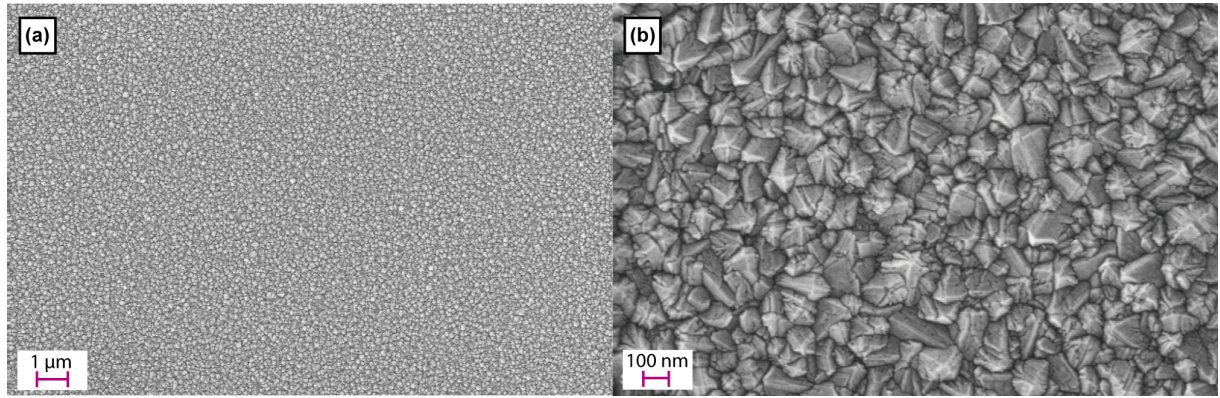

*Figure 3. SEM images of a 1µm-thick diamond film grown onto a silicon substrate in the NIRIM reactor. (a) Lower SEM magnification to show the coalescence of the diamond film and (b) a higher SEM magnification showing 100 nm diamond crystalline grains.*

To confirm the quality of the films grown, Raman spectroscopy measurements were carried out with an inVia Renishaw confocal Raman microscope equipped with a 532 nm laser. Figure 4 plots the normalized Raman intensity from 1000 cm$^{-1}$ to 1650 cm$^{-1}$ of the sample after background subtraction. The spectra shows a clear first-order diamond peak at 1332 cm$^{-1}$ [24,25], broadened due to the small size of the crystallites as characteristic of NCD films[26]. Less pronounced is the G-band peak, located around 1550 cm$^{-1}$ and attributable to the in-plane stretching of sp$^2$ bonds within graphitic-like material[27]. Finally, the peaks at 1130 cm$^{-1}$ and at 1450 cm$^{-1}$ arise from the presence of C=C and C-H bonds within transpolyacetylene (TPA) chains at grain boundaries between neighbouring crystallites[28]. With the scattering efficiency of graphite previously reported to be ~50 times larger than that of diamond at an excitation of 514 nm, in combination with the morphology observed with SEM, it can therefore be concluded that the diamond fraction within the NCD film is high[29].

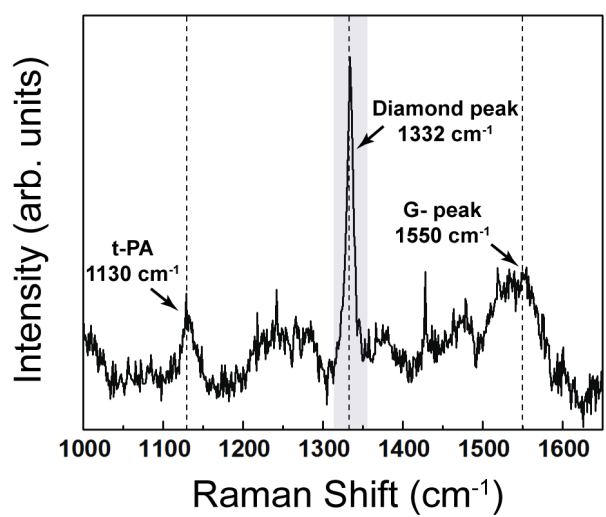

*Figure 4. Raman spectrum of the diamond film. The peak at 1332 cm$^{-1}$ is attributed to the first-order diamond peak and the band centred at 1550 cm$^{-1}$ is known as the G-band, attributable to the in-plane stretching of graphitic like structures.*

**B. Reactor cleaning**

After approximately 20 hours of operation at the conditions mentioned above a thin band of amorphous carbon built up around the interior of the quartz tube immediately adjacent to the plasma. To prevent the conductive lacquer from affecting the propagation of microwaves through the waveguide the quartz tube was then removed, placed within a furnace, and heated in air at 600-700 °C until clean. The tube was then reused for subsequent intrinsic growth. Alternatively, the deposits can also be removed through abrading with tissue paper and solvent. The quartz rod meanwhile remained relatively clean as a result of being enshrouded within the sample holder within the plasma, while the holder itself was cleaned with isopropyl alcohol and DI after each run.

**IV. CONCLUSIONS**

The design of a simple and space constrained NIRIM type CVD reactor has been laid out, enabling replication as an alternative to commercial diamond plasma deposition systems. The reactor consists of predominantly off-the-shelf fittings and utilises a modular construction to allow for simple modification, repair and cleaning, while also ensuring ease of use during growth runs. SEM images of NCD films grown with the reactor show high crystallinity while Raman spectra indicate a significant diamond fraction with minimal non-diamond content, demonstrating the efficacy of the reactor design.

ACKNOWLEDGMENTS

The authors thank the Royal Society International Exchanges Scheme (IE131713) and EU FP7 FET Open "Wavelength tunable Advanced Single Photon Sources". The authors also acknowledge Sairem for providing CAD files of the microwave components used.